\def\lte{\lower 0.5ex\hbox{${}\buildrel<\over\sim{}$}}
\def\gte{\lower 0.5ex\hbox{${}\buildrel>\over\sim{}$}}

\def\gp{\gamma_+} \def\gm{\gamma_-} \def\bp{\beta_+} \def\bm{\beta_-}
\def\gc{\gamma_{cm}} \def\gem{\gamma_{1-}} \def\gzm{\gamma_{2-}}
\def\gep{\gamma_{1+}} \def\gzp{\gamma_{2+}} \def\gpmi{\gamma_{\pm i}}
\def\gepmi{\gamma_{1\pm i}} \def\gzpmi{\gamma_{2\pm i}}
\def\gu{\gamma_{\infty}} \def\kp{\kappa_+} \def\km{\kappa_-}
\def\vdsw{\overline{v {d\sigma \over dk}}} \def\vdsv{\left< \vdsw
\right>} \def\vdswr{\left(\vdsw \right )_{rel}} \def\vdsvr{\vdsv_{rel}}
 \def\mkgem{\max\{k, \gem\}} \def\mkgep{\max\{k,
\gep\}} \def\mkgzm{\min\{k, \gzm\}} \def\mkgzp{\min\{k, \gzp\}}
\def\dns{\delta_{s, N}} \def\bmv{\overrightarrow\beta_-}
\def\bpv{\overrightarrow\beta_+} \def\bcmv{\overrightarrow\beta_{cm}}
 
\def\bg{\beta_{\Gamma}}  \def\Npm{N^{\pm}}
   
\def\gpm{\gamma_{\pm}} \def\gmp{\gamma_{\mp}} \def\gepm{\gamma_{1\pm}}
\def\gemp{\gamma_{1\mp}} \def\gzpm{\gamma_{2\pm}}
\def\bg{\beta_{\Gamma}} \def\N{I \!\!  N}

\def\abb{\par\noindent \hangindent=5.7em \hangafter=1}

\documentstyle[epsf, rotate]{l-aa}
\begin{document}

\thesaurus{02.02.1 -- 02.18.5 -- 11.01.2 -- 11.14.1 -- 13.07.3}
\title{Pair annihilation radiation from relativistic jets in $\gamma
$-ray blazars} \author{M.  B\"ottcher, R. Schlickeiser}
\institute{Max-Planck-Institut f\"ur Radioastronomie, Postfach 20 24, 53
010 Bonn, Germany}
\date{Received ; accepted 06.07.1995}
\offprints{M. B\"ottcher}

\maketitle

\begin{abstract} The contribution of the pair annihilation process in
relativistic electron-positron jets to the $\gamma -$ray emission of
blazars is calculated.  Under the same assumptions as for the
calculation of the yield of the inverse Compton scattered photons by
relativistic particles in the jet (Dermer and Schlickeiser 1993) we
calculate the emerging pair annihilation radiation taking into account
all spectral broadening effects due to the energy spectra of the
annihilating particles and the bulk motion of the jet.  It is shown that
in spectral appearance the time-integrated pair annihilation spectrum
appears almost like the well-known $\gamma $-ray spectrum from decaying
$\pi ^o$-mesons at rest, since it yields a broad bumpy feature located
between 50 and 100 MeV. We also demonstrate that for pair densities
greater $10^9$ cm$^{-3}$ in the jet the annihilation radiation will
dominate the inverse Compton radiation, and due to its bump-like
spectrum indeed may explain reported spectral bumps at MeV energies.
The refined treatment of the inverse Compton radiation including
Klein-Nishina cross section corrections, the low-energy cutoff in the
radiating relativistic particle distribution function and a realistic
greybody accretion disk target photon spectrum leads to spectral
breaks of the inverse Compton emission in the MeV energy range with a change
in spectral index $\Delta \alpha $ larger than 0.5 as detected in PKS~0528+134
and 3C273.

 \end{abstract}

\section{Introduction}

Until now 45 extragalactic $\gamma $-ray sources (status October 1994)
have been clearly detected by the EGRET experiment aboard the Compton
Observatory (Kanbach 1994, Thompson 1994).  Several of them are also
seen by the COMPTEL and OSSE instruments (Kurfess 1994). The broadband
high energy spectra extend over four decades in energy and exhibit quite
pronounced spectral forms and breaks.  Apart from the Large Magellanic
Cloud all extragalactic $\gamma $-ray sources are radio-loud, have a
compact core and are known as highly variable sources. A large
fraction of these sources exhibit apparent superluminal motion.

Because of its rapid variability, its high compactness, lack of
radio-quiet sources in the sample and high presence of superluminal
motion sources it is generally agreed (Blandford 1993, Dermer 1993,
Dermer and Schlickeiser 1992, Dermer \& Gehrels 1995) that the $\gamma
$-ray emission originates in strongly beamed sources, in accord with the
relativistic beaming hypothesis that has served well as the baseline
model for the central engines of active galactic nuclei (Scheuer and
Readhead 1979, Blandford and K\"onigl 1979).  Dermer and Schlickeiser
(1993) - hereinafter referred to as DS - have demonstrated that at hard
X-ray and $\gamma $-ray frequencies inverse Compton scattering of
accretion disk photons by relativistic electrons and positrons in an
outflow ing plasma with relativistic bulk motion (bulk Lorentz factor
$\Gamma \simeq 5-10$) dominates over other radiation mechanisms.  This
$\gamma $-ray production process succesfully explains the key
observational features as (i) the peak of the bolometric luminosity at
$\gamma $-ray frequencies (ii) the $\gamma $-ray variability time scale
$\le $ few days, (iii) the softening of the spectra between the hard
X-ray and medium $\gamma $-ray regime, (iv) the high percentage of
superluminal sources in the sample, (v) the generation of TeV emission
for favourably aligned observers.

It has been suggested (Henri, Pelletier \& Roland 1993) that pair
annihilation in relativistic electron-positron jets could be of
considerable importance to such jet models.  The purpose of this study
is thus to extend the DS-model by including the $\gamma $-ray production
from pair annihilation radiation from relativistic jets. Under the same
assumptions as for the calculation of the inverse Compton scattering
from the jet we will determine the yield of annihilation radiation.  Of
special interest will be the modelling of spectral breaks with a change
in spectral index $\Delta \alpha $ larger than 0.5 as detected in PKS~0528+134,
$\Delta \alpha =1.16 (+0.23,-0.11)$ and 3C273, $\Delta \alpha
=0.66 (+0.13,-0.15)$.  In the original DS approach spectral breaks up
to $\Delta \alpha \le 0.5$ were obtained by incomplete inverse Compton
cooling of an injected power law during the time evolution of the
$\gamma $-ray flare. Taking into account all spectral broadening
effects due to the energy spectra of the annihilating particles and the
bulk motion of the jet, we show here that the contribution from this
radiation component peaks near a few MeV, and if it dominates the
inverse Compton component it may provide an explanation for the reported
spectral maxima in this energy range.

\section{Jet model for the $\gamma $-ray emission}

The jet model for the $\gamma $-ray emission in AGNs of DS is
illustrated in Figure 1. Relativistic electrons and positrons, with
isotropic energy distribution functions $n_{\pm }(\gamma ;z)$ in the
comoving frame at height $z$ above the accretion disk, f low outward
along the disk symmetry axis with bulk Lorentz factor $\Gamma $ and bulk
velocity $\beta_{\Gamma }c$ from an initial height $z_i$.  Photons
produced by the accretion disk pass through the jet and are Compton
upscattered to $\gamma $-ray energi es $km_ec^2$ (see DS for details).
The relativistic electrons and positrons fill not the whole jet but only
an (assumed) spherical blob of radius $R_B$.  We follow the notation of
DS: energies and angles without asterisks denote values in the comoving
fluid frame, while these quantities with asterisks denote values in the
stationary frame $K$ of the accretion disk. We do not include cosmological
effects here.

\begin{figure} \epsfxsize=6cm \rotate[r]{\epsffile [15 5 350 220]
{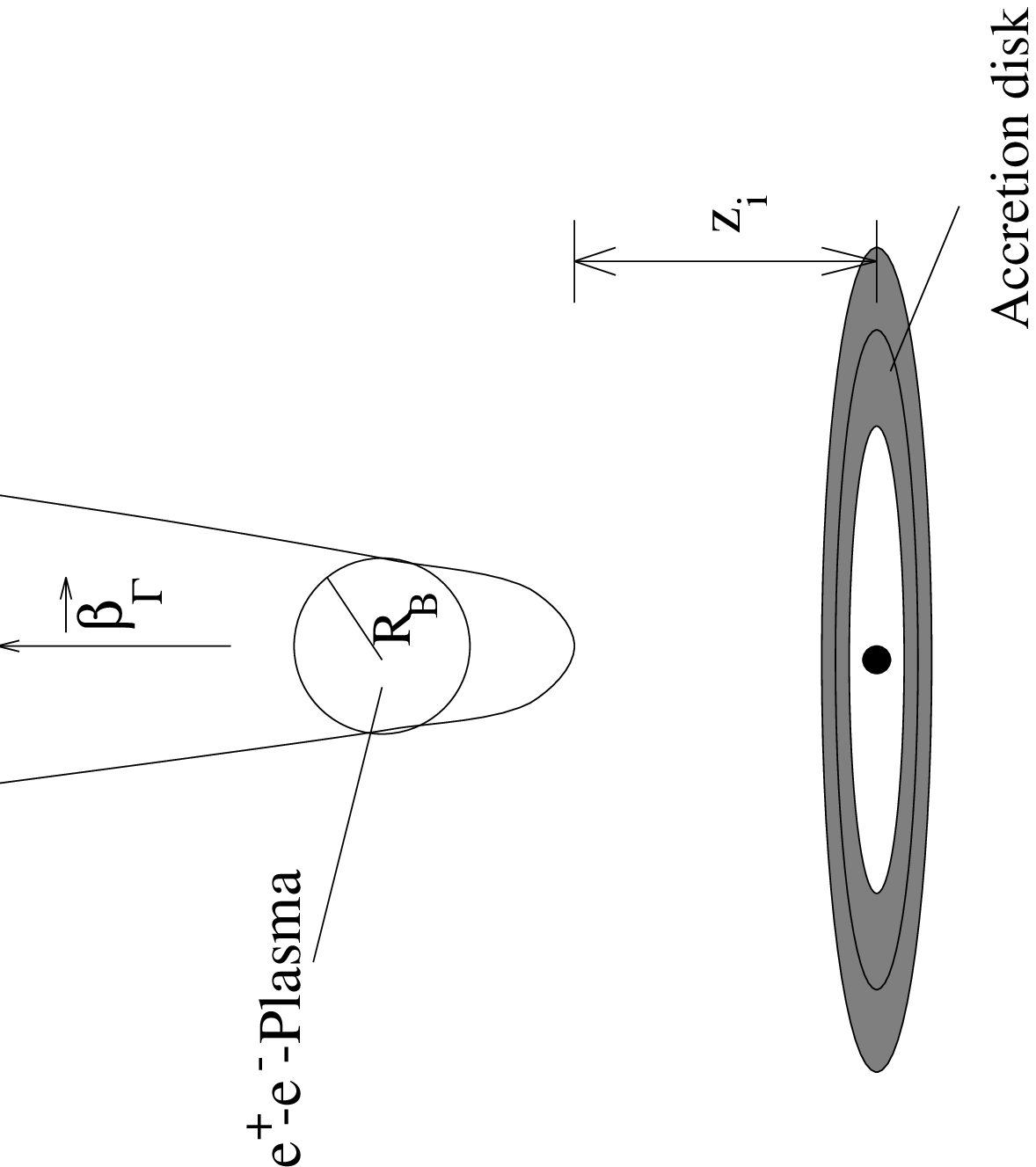}} \abb {\bf Fig. 1:  } Jet model for the $\gamma$-ray
emission in AGNs \bigskip \end{figure}

To avoid heavy attenuation of the produced GeV $\gamma $-rays by pair
production interactions with the accretion disk photons Dermer and
Schlickeiser (1994) have pointed out that the initial height of the blob
has to be larger than $z_p= 10^{15} (L_{46}/\Theta _{50})^{1/2} $cm
where $L_{46}$ denotes the accretion disk luminosity in units of
$10^{46} $ erg/s and $\Theta _{50}$ the accretion disk temperature in
units of $50 $ eV.  Such a high location of the blob is also consistent
with the presence of many superluminal motion sources in the sample of
$\gamma $-ray blazars. At these heights the pair annihilation radiation
freely escapes from the system and does not initiate a pair cascade
(Lightman and Zdziarski 1987, Zdziarski et al. 1990, Protheroe et al.
1992) with plenty production of 511 keV annihilation line radiation from
the cascade-cooled cold electron-positron pairs that is not observed
from these objects (Kurfess 1994). As in DS we start wit h initial
relativistic power law energy spectra for the radiating particles

$$ n_{\pm }(\gamma _{\pm },z_i)= N_{\pm }\gpm^{-s}\; \; \; \; \; 1 \ll
\gepm \le \gpm \le \gzpm, \eqno (1)$$

and follow the time evolution of the energy spectra under the various
energy loss processes. These spectra are used to calculate the yield of
annihilation radiation at various heights. We start our analysis with
the calculation of the annihilation radiation emissivity.

\section{Pair annihilation radiation from the electron-positron blob}

The general calculation of the yield of pair annihilation radiation from
relativistic plasmas has been treated by Svensson (1982). In the case
of an ultrarelativistic pair plasma, characterised by Eq. (1), this
treatment can be considerably simplified by using the differential pair
annihilation cross section

$$ \left({d \sigma \over d\Omega_{cm}}\right)_{rel} $$ $$ = {r_e^2 \over
4 \gc^2} \left\lbrace -1 \> + \> \left( {1 \over 1 - \beta_{cm}x} + {1
\over 1 + \beta_{cm} x}\right) \right\rbrace \eqno(2) $$

which is valid for $\gamma_{cm} \gg 1$.  The subscript $cm$ denotes
quantities measured in the center-of-momentum frame and $x =
{\overrightarrow k_{cm} \bcmv \over k_{cm} \beta_{cm}}$.  Then,
following the calculation leading to Svensson's formulae (55) -- (57)
yields

$$ \left({d\sigma \over dk}\right)_{rel} = \pi \, r_e^2 \> \Biggl(- {1
\over \gc^2 \, \sqrt{(\gp + \gm)^2 - 4 \gc^2}} \> + $$ $$ + \> {2 \over
\gc \, \sqrt{ \gc^{*2} + c_+ \gc^2}} \> + \> {2 \over \gc \, \sqrt{
\gc^{*2} + c_- \gc^2}} \Biggr) \eqno(3) $$

--- where

$$ c_{\pm} = (k - \gmp)^2 \, - \, 1, \eqno(4) $$ $$ \gc^{*2} = k \> (\gp
+ \gm - k), \eqno(5) $$

k is the photon energy normalized to the electron rest energy, the
subscript $\pm$ denotes quantities refering to the motion of the
positron/electron as measured in the labor frame (the rest frame of the
plasma blob) and $\gc^*$ is the maximum center-of-momentum
$\gamma$-factor in the case of which a photon of energy $k$ can be
produced, which is the notation used by Svensson (1982) --- and

$$ \vdswr (k, \gp, \gm) = {c \, \pi \, r_e^2 \over \gp^2 \, \gm^2} \cdot
$$ $$ \cdot \left( \sqrt{(\gp + \gm)^2 - 4 \gc^2} \, + \, H_+ \, + \,
H_- \right) \Biggr\vert_{\gc^L}^{\gc^U} \eqno(6) $$

where the bar denotes the averaging over collision angles between $\bpv$
and $\bmv$ and

$$ H_{\pm} = {\gc \over c_{\pm}} \sqrt{\gc^{*2} + c_{\pm} \gc^2} \> + $$
$$ + \> {\gc^{*2} \over {\left\vert c_{\pm} \right\vert}^{3/2}} \cdot
\cases{\arcsin\left ( {\sqrt{-c_{\pm}} \> \gc \over \gc^*}\right ) & for
$c_{\pm} < 0$ \cr -$arsinh$\left ({\sqrt{c_{\pm}} \> \gc \over
\gc^*}\right ) & for $c_{\pm} > 0$ \cr}, \eqno(7) $$ \medskip

$$ H_{\pm} = {2 \over 3} {\gc^3 \over \gc^*} \hskip 2cm \hbox{for} \;\>
c_{\pm} = 0 \eqno(8) $$

Also the integration limits $\gc^U$, $\gc^L$ which are determined by
kinematical restrictions may be simplified.  As given by Svensson (1982)
these integration limits are found to be

$$\gc^L = \gc^{min} = \sqrt{{1 \over 2} \left( 1 + \gp \gm [ 1 - \bp \bm]
\right)} \eqno(9 a) $$
$$\gc^U = \min \, \{\gc^{max}, \gc^*\} \eqno(9 b) $$

where

$$ \gc^{max} = \sqrt{{1 \over 2} \left(1 + \gp\gm \left[1 +
\bp\bm\right]\right)} \eqno(10) $$

is the center-of-momentum $\gamma$-factor in the case of head-on
collision which, for $\bp, \, \bm \approx 1$, simplifies to $\gc^{max}
\approx \sqrt{\gp \gm}$. $\gc^{min}$ is the minimum center-of-momentum
Lorentz factor in the case of an overtaking collision. Writing
$\gc^{*2} = \gp\gm - (k - \gm)(k -\gp)$, we immediately see that

$$ \gc^U = $$ $$ \cases{ \gc^{max} & for $\; (k > \gp \wedge k < \gm)\;
$or$ \; (k < \gp \wedge k > \gm)\; $ \cr \gc^* & for $\; (k < \gp \wedge
k < \gm) \; $or$ \; (k > \gp \wedge k > \gm) \; $ \cr} \eqno(11) $$

Furthermore, we have to satisfy the condition

$$ k < k_U $$ $$ = {1 \over 2} \left( \gp [1 + \bp] + \gm [1 +
\bm]\right) \eqno(12) $$

where $k_U$ is the maximum photon energy available by the annihilation
of an $e^+$-$e^-$-pair characterized by $\gp, \, \gm$ (Svensson 1982).
In the ultrarelativistic case, equation (12) simply becomes $k < \gp +
\gm$ which is equivalent to the condition $\gc^{*2} > 0$.

Unfortunately, equation (6) is still not analytically integrable, but
for $\gep, \, \gem \, \gte 5$, it may be very well used for numerical
calculations of the integral

$$ \vdsv (k) $$ $$ = \int\limits_{\gep}^{\gzp} d\gp
\int\limits_{\gem}^{\gzm} d\gm \> f_+(\gp) \, f_-(\gm) \> \vdsw (k, \gp,
\gm) \eqno(13) $$

(Svensson 1982) where the brackets denote the averaging over the
distribution functions of electrons and positrons.

For further simplifications, we use the fact that in Eq.  (6) only the
upper limit $\gc^U$ is important and also the first term is negligible
as compared to the values of $H_{\pm} (k, \gc^U, \gc^*)$.  As mentioned
by Svensson, for highly relativistic pair p lasmas, the angle averaged
emissivity is sharply peaked at $\gp = k$ and $\gm = k$ where the photon
takes over the energy and momentum of the incoming positron or electron,
respectively. We use this fact to approximate $\vdsw$ by constructing a
function that satisfies the condition of these peak values and exhibits
the same asymptotical behavior as found for $\vdsw$.

The well-known $\delta$-function approximation does not work well in the
case of an energy distribution as given by eq.  (1), particularly, it
does not describe the lower energy tail of the annihilation spectrum ($k
< \min \, \{\gep, \gem \}$) at all.  The approximation scheme we derive
here is much simpler than that given by Coppi and Blandford (1990) who
still need numerical integrations to determine the dispersion of $\vdsw$
about $k$ but whose results are --- in contrast to ours --- also valid
for midly relativistic pair plasmas.

The peak values of $H_{\pm}$ as given by eq.  (6) are easily found by
setting $c_{\pm} = -1$ and $\gc^U = \gc^* = \sqrt{\gp\gm}$:

$$ H_{\pm} (k, \gc^U) \biggr\vert_{c_{\pm} = -1} = {\pi \over 2}
\left(\gc^U\right)^2.  \eqno(14) $$

Considering the asymptotic behavior of $\vdsw$, we see that for $\vert k
- \gpm \vert \gg 1$ the photon is not emitted in the direction of the
incoming electron/positron, and so $x$ does not approach $1$ where the
differential cross section (2) peaks.  For this reason, we may set
$\beta_{cm} = 1$ in eq.  (2) and in the following steps of calculation,
which yields

$$ \vdsw (k, \gp, \gm) \biggr\vert_{c_{\pm} \gg 1} = {\left( \gc^u
\right)^2 \over \vert k - \gmp \vert }. \eqno(15) $$

A quite simple function combining the characteristics described by eqs.
(14) and (15) and therefore approximating $\vdsw$ is

$$ \vdswr (k, \gp, \gm ) $$ $$ \approx {c \pi r_e^2 \over \gp^2 \gm^2}
\cdot \left[ {\left ( \gc^U \right )^2 \over \vert k - \gp \vert + {2
\over \pi} } + {\left ( \gc^U \right )^2 \over \vert k - \gm \vert + {2
\over \pi} } \right ]. \eqno(16) $$

\begin{figure} \epsfxsize=6cm \epsffile [30 180 430 560] {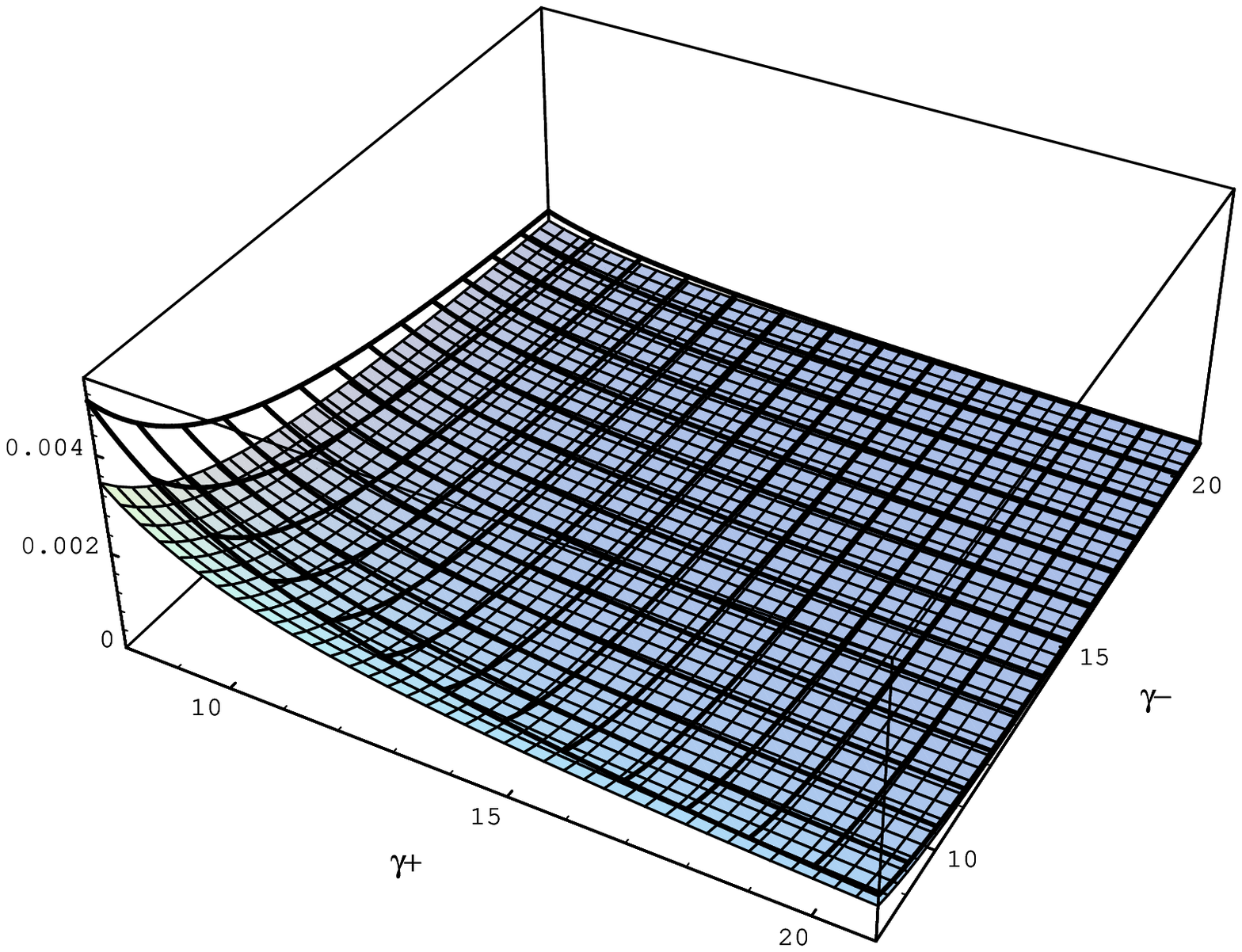} \abb
{\bf Fig. 2:  } Asymptotic approximation (16) (transparent area) in
comparison to the relativistical value of $\vdsw$ calculated by eq.  (6)
(gray area) for $k = 4$ \bigskip \end{figure}

\begin{figure} \epsfxsize=6cm \epsffile [30 180 430 560] {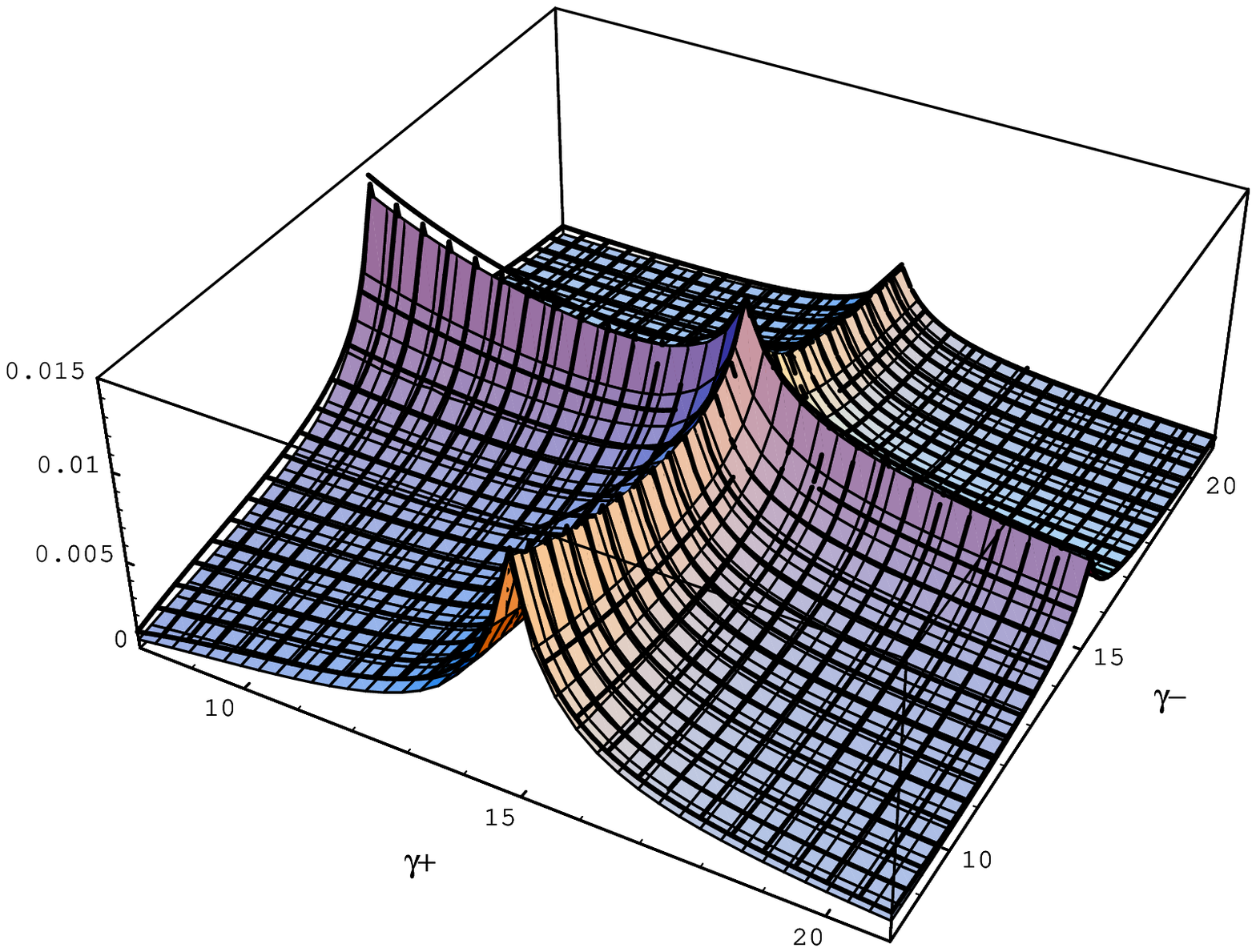} \abb
{\bf Fig. 3:  } Same as Fig. 2, but for $k = 15$ \bigskip \end{figure}

\begin{figure} \epsfxsize=6cm \epsffile [30 180 430 560] {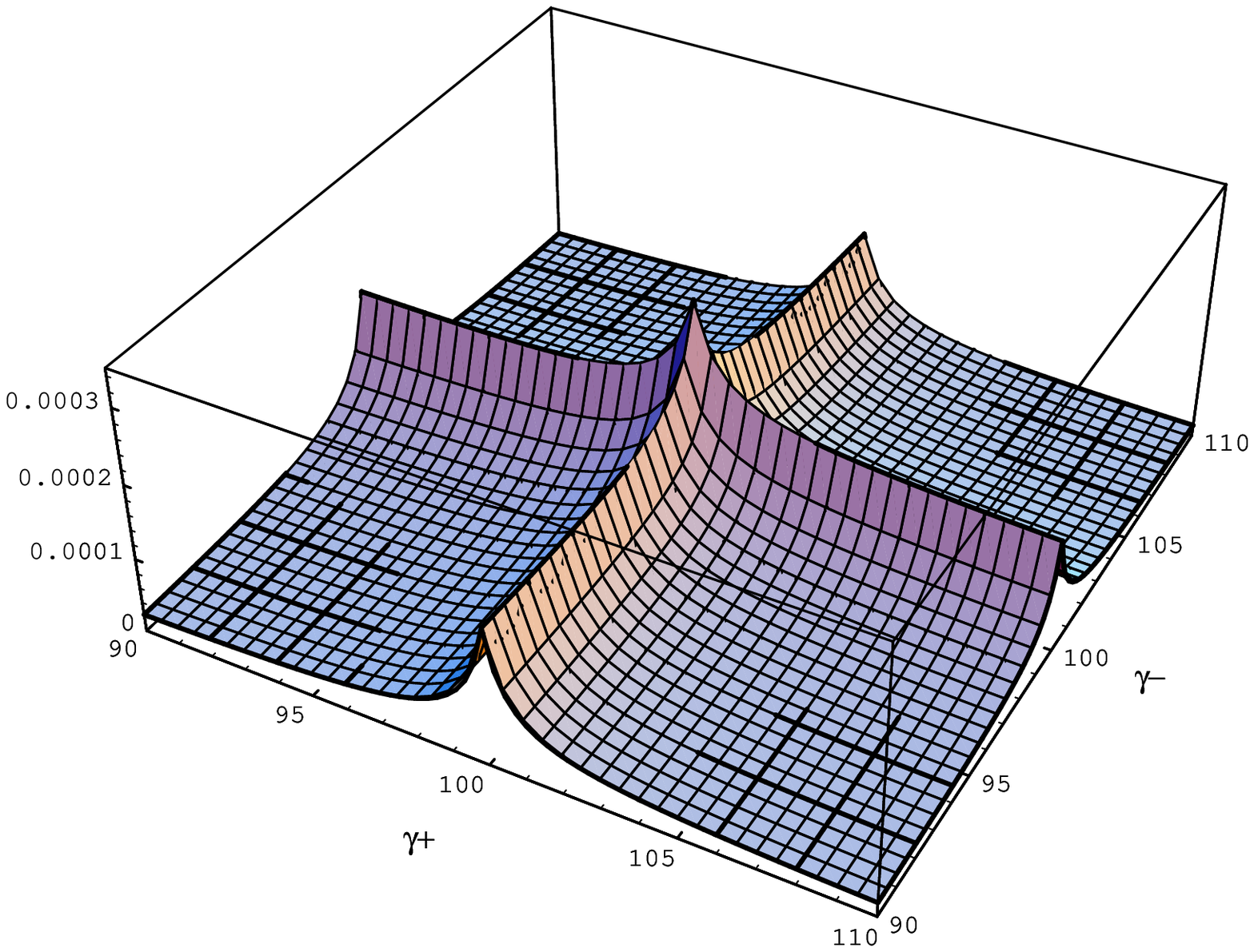} \abb
{\bf Fig. 4:  } Same as Figs. 2 and 3 for $k = 100$ \bigskip
\end{figure}

We call it the asymptotic approximation.  Figures 2 -- 4 show that Eq.
(16) is a very useful approximation to the angle averaged emissivity
from highly relativistic pair plasmas for $k \gte \min\, \{ \gep, \gem
\}$, but which is --- in contrast to the $\delta$-function
approximation --- still sensible for $k < \min\, \{ \gep, \gem \}$.

By means of a series expansion of the asymptotic approximation, the
integrations in eq.  (13) are elementary for power-law distributions (1)
and are also analytically executable for other physically relevant
ultrarelativistic distribution functions.  Due to different
determinations of $\vdsw$ in the four integration areas defined by Eq.
(11), the results look somewhat involved, but they are --- after some
manipulations --- numerically quite simply evaluable.

Expanding (16) in a series and performing the integral (13) --- where we
neglect the integration area given by the edge points $(\gpm, \gmp) =
(k, k - \gepm), (k, \gepm), (\gepm, k - \gemp), (\gemp, k - \gepm),
(\gep, k)$ for integration over ${1 \over {2 \over \pi} + \vert k -
\gpm}$ --- yields

$$ \vdsvr (k) = $$ $$ \Theta (k - \gem) \Theta (k - \gep)
\biggl({\gem^{-s} - \left( \mkgzm \right )^{-s} \over s} \> \cdot $$ $$
\cdot \> \bigl[ S_1(\mkgzp , s+1) - S_1(\max\{\gep, k - \gem\} , s+1)
\bigr] $$ $$ + \; {\gem^{-(s+1)} - \left ( \mkgzm \right )^{-(s+1)}
\over s + 1} \> \cdot $$ $$ \cdot \> \bigl[ S_1(\mkgzp , s) -
S_1(\max\{\gep, k - \gem\}, s) $$ $$- \; k \lbrace S_1(\mkgzp , s+1) $$
$$ - S_1 (\max\{\gep, k - \gem\}, s+1) \rbrace \bigr] \biggr) \eqno(17
a) $$ \smallskip $$ + \; \Theta (\gzm - k) \Theta (k - \gep )
{\left(\mkgem\right)^{-s} - \gzm^{-s} \over k \; s } \> \cdot $$ $$
\cdot \> \bigl( S_1(\mkgzp , s) - S_1(\gep, s) \bigr) \eqno(17 b) $$
\smallskip $$ + \; \Theta (k - \gem ) \Theta ( \gzp - k){\left
(\mkgzm\right )^{-s} - \gem^{-s} \over s} \> \cdot $$ $$ \cdot \> \bigl(
S_2(\gzp , s+1) - S_2(\mkgep , s+1) \bigr) \eqno(17 c) $$ \smallskip $$
+ \; \Theta ( \gzm - k) \Theta ( \gzp - k) \biggl({\gzm^{-s} -
\left(\mkgem \right)^{-s} \over s } \> \cdot $$ $$ \cdot \, k \, \bigl(
S_2(\gzp , s + 2) - S_2(\mkgep , s + 2) \bigr) $$ $$ + \; {\gzm^{-(s+1)}
- \left(\mkgem\right)^{-(s+1)} \over s + 1} \; k \cdot $$ $$ \cdot
\Bigl(S_2(\gzp , s+1) - S_2(\mkgep , s+1) $$ $$ - k \bigl[S_2(\gzp ,
s+2) - S_2(\mkgep , s+2) \bigr] \Bigr) \biggr) \eqno(17 d) $$ $$ + \;
\bigl[ + \leftrightarrow - \bigr] $$ \medskip

where $[ + \leftrightarrow - ]$ denotes the preceding terms (17 a) --
(17 d) under transposition of the subscripts $+ \leftrightarrow -$,
$\Theta$ is the Heaviside function and

\medskip $$ S_1(\gamma , s) = \gamma^{-s} \sum\limits_{\buildrel {n = 0}
\over {n \ne s} }^{\infty} {1 \over n - s} \left({\gamma \over
\kp}\right)^n \; + \; \dns \> k^{-s} \ln (\gamma ) \eqno(18 a) $$ $$
S_2(\gamma , s) = \gamma^{-s} \sum\limits_{n = 0}^{\infty} {1 \over n +
s} \left({\km \over \gamma}\right)^n, \eqno(18 b) $$

with the symbol $\dns$ defined by

$$ \dns = \cases{ 1 & for $s \in \N$ \cr 0 & for $s \notin \N$ \cr }
\eqno(19) $$

and $\kappa_{\pm} = k \pm {2 \over \pi}$.  Of course, we must not
replace $\kappa_{\pm}$ by $k$ in eqs.  (18), because the series $S_1$
and $S_2$ would diverge then.  The series $S_1$ is easily evaluable by
summing several ($> s$) terms and replacing the rem aining part of the
sum by the integral whose upper Riemann sum it is, while $S_2$ may be
evaluated by subtracting a geometrical series whose value is known, with
the remaining series converging very quickly.

\begin{figure}
\epsfxsize=6cm
\epsffile [30 60 410 680] {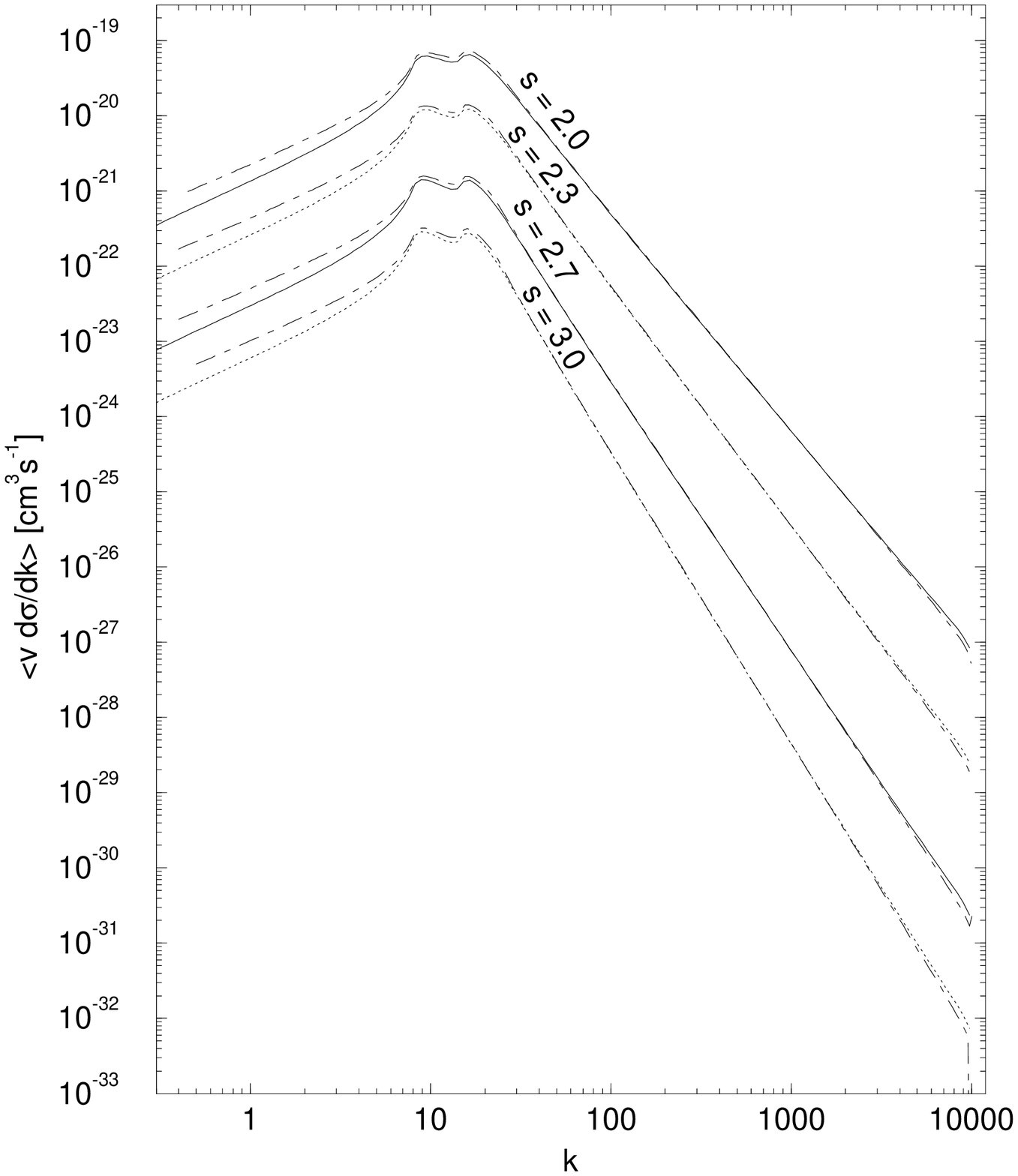} \par
{\bf Fig. 5:} Pair annihilation spectra calculated by the asymptotic
approximation (17) (dashed) compared to numerical integrations (solid
and dotted, respectively); parameters:  ${\gep = 8},$ ${\gem = 15},$
$\gzm = \gzp = 10^4$ \bigskip \end{figure}

Doing so, we see that Eq. (17) provides a very accurate approximation
to the pair annihilation spectrum.  In Fig. 5, this approximation is
compared to the results of numerical integrations over the exact
solution for $\vdsw$ given by Svensson (1982).  Especially the peak
structure and the high energy tail of the spectrum are very well
described by the asymptotic approximation, while also the low energy
tail is sensibly approximated, considering that this part of the
spectrum --- as compared to spectra produ ced by other radiation
mechanisms resulting mainly in power laws --- will not be very
important.  Note that the different absolute values of the peak
emissivity for different spectral indices is caused by the fact that for
simplicity we used not normalized distribution functions with $N_{\pm} =
1$.

\section{Particle spectra evolution}

We now consider the evolution of an ensemble of particles that is
injected at height $z_i$ with a power law distribution in energy given
by Eq. (1).  We first treat the effects of inverse-Compton, synchrotron
bremsstrahlung and pair annihilation losses seperately in order to
achieve an estimate for typical timescales for the deviation from an
initial power law distribution caused by each one of these processes.

The typical peak structure of annihilation radiation is mainly caused by
the lower energy end of particle distributions.  Therefore we show that
energy loss mechanisms most effectivly cooling high energy particles
($\gpm \gg \gepm$) may not seriously influence the shape of the
annihilation spectrum during the evolution of the particle spectrum.
So, the typical timescale for low energy particle spectra evolution may
be significantly longer than the timescale for energy losses by a single
particle.

Note that this does not represent a realistic calculation of the
particle spectra evolution --- this will be treated in future work ---,
but gives only a rough estimate of the typical timescale on which the
assumption of a power law distribution of particle energies is
justified.

\subsection{Inverse Compton losses}

The effect of inverse Compton losses on particle energy spectra in
the jet has been treated in DS.  There, a typical particle energy $\gu$
is defined by

$$ \gu = {\tilde z_i \over 2488 \> \ell_0} \; \bg \, \Gamma^3 \, (1 +
\bg)^2 \eqno(20) $$

where the tilde denotes the normalization to gravitational radii ---
$\tilde z = {z \over R_g}$, $R_g = {GM \over c^2}$ --- and $\ell_0 =
{L_0 \over L_{edd}}$.  It is shown that particles with energy greater
than $\gu$ are suffering heavy losses while the spectral shape of
particles with energy $\gpm < \gu$ remains nearly unaffected by inverse
Compton losses.  This spectral evolution which, of course, does not
depend on the particles' charge is shown in Fig. 6.

\begin{figure}
\epsfxsize=6cm
\epsffile [30 60 410 680] {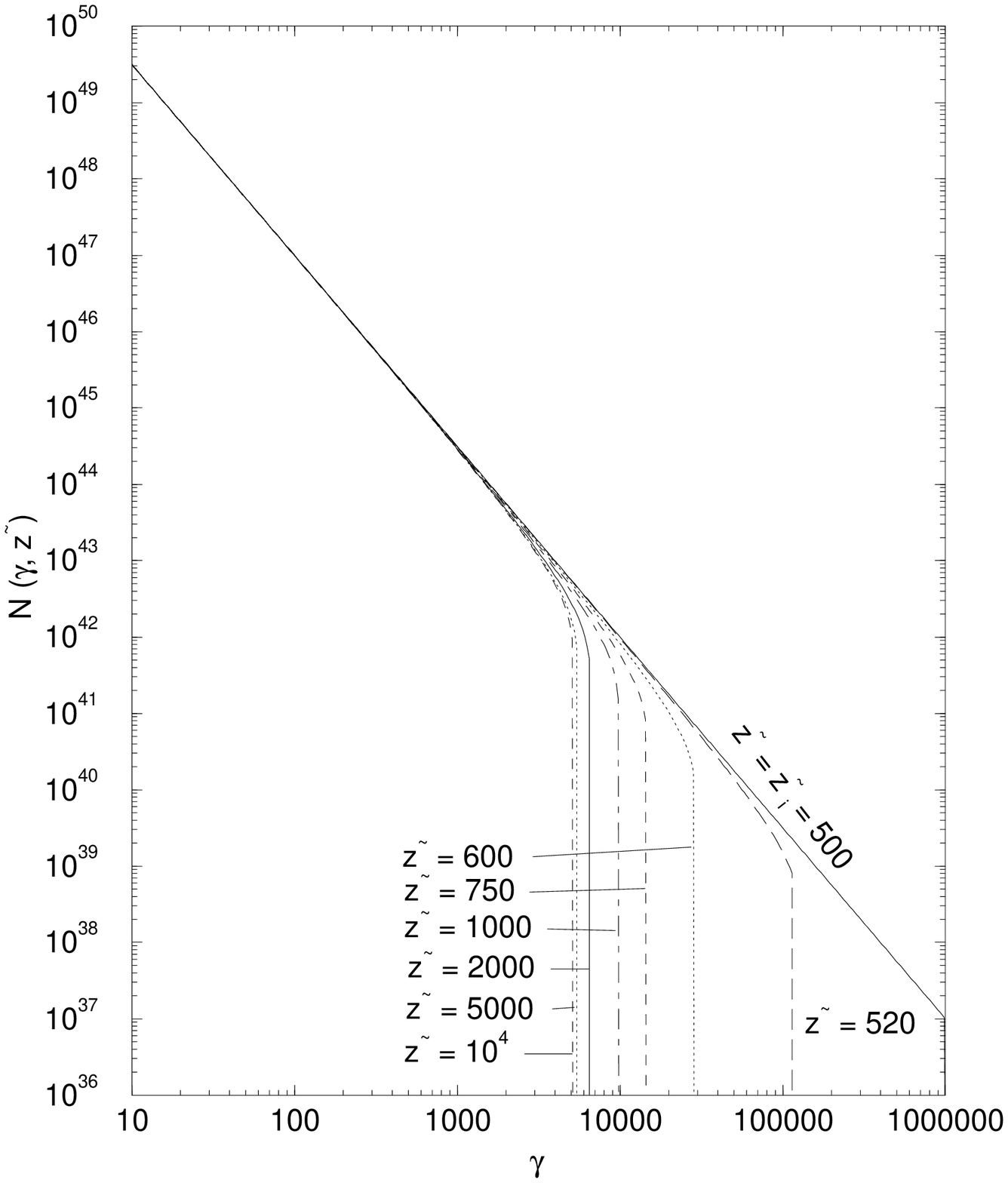} \par
{\bf Fig. 6:  } Evolution of a power law particle spectrum, affected by
inverse-Compton losses.  Parameters:  $\tilde z_i = 500$, $\gepmi = 10$,
$\gzpmi = 10^6$, $s = 2.5$, $M_8 = 1$, $\ell_0 = 0.1$.  \bigskip
\end{figure}

{}From Eq.  (20) and Fig. 6, we conclude that the shape of the low energy
particle spectrum will not be altered substantially by inverse Compton
losses, and therefore the pair annihilation peak structure will not
strongly depend on particle spectra evolution caused by the inverse
Compton process.

\subsection{Synchrotron losses}

In order to investigate the effect of synchrotron losses we use the
single particle energy loss rate due to the synchrotron process

$$ - m c^2 \> \dot \gpm \> = \> {4 \over 3} \> \sigma_T \, {B^2 \over 8
\pi} \> c \> \beta_{\pm}^2 \gamma_{\pm}^2 \eqno(21) $$

--- where B is the magnetic field strength which we assume to be
approximately constant --- combined with the blob's equation of motion

$$ \tilde z (t^*) = \tilde {z_i} + {c \bg t^* \over R_g} \eqno(22) $$

and the time-dilation $\delta t^* = \Gamma \delta t$.  This yields

$$ \left( {d \gpm \over d \tilde z} \right)_{Sy} = - {4 \over 3} \> {
\sigma_T \, B^2 \over 8 \pi m c^2} \> { R_g \over \Gamma \, \bg} \>
\gpm^2 \> =:  \> -K \> \gpm^2.  \eqno(23) $$

where

$$ K = 5 \cdot 10^{-7} \> {M_8 \over \Gamma} \> \left( {B \over G}
\right)^2.  \eqno(24) $$

The differential equation (23) is solved by

$$ \gpm^{-1} (\tilde z) = \gpmi^{-1} \, + \, K \, (\tilde z - \tilde
z_i) \eqno(24) $$

which is valid in the bounds

$$ \gepm := \left( \gepmi^{-1} + K [\tilde z - \tilde z_i] \right)^{-1}
\le \gpm $$ $$ \gpm \le \left( \gzpmi^{-1} + K [\tilde z - \tilde z_i]
\right)^{-1} =:  \gzpm.  \eqno(25) $$

Considering conservation of particle number and inserting eq.  (1) for
the initial state leads to the resulting spectrum

$$ n_{\pm} (\gpm, \tilde z) = N_{\pm} \gpm^{-2} \left( \gpm^{-1} \, - \,
K \, [\tilde z - \tilde z_i] \right)^{s-2} \eqno(26) $$

\begin{figure}
\epsfxsize=6cm
\epsffile [30 60 410 680] {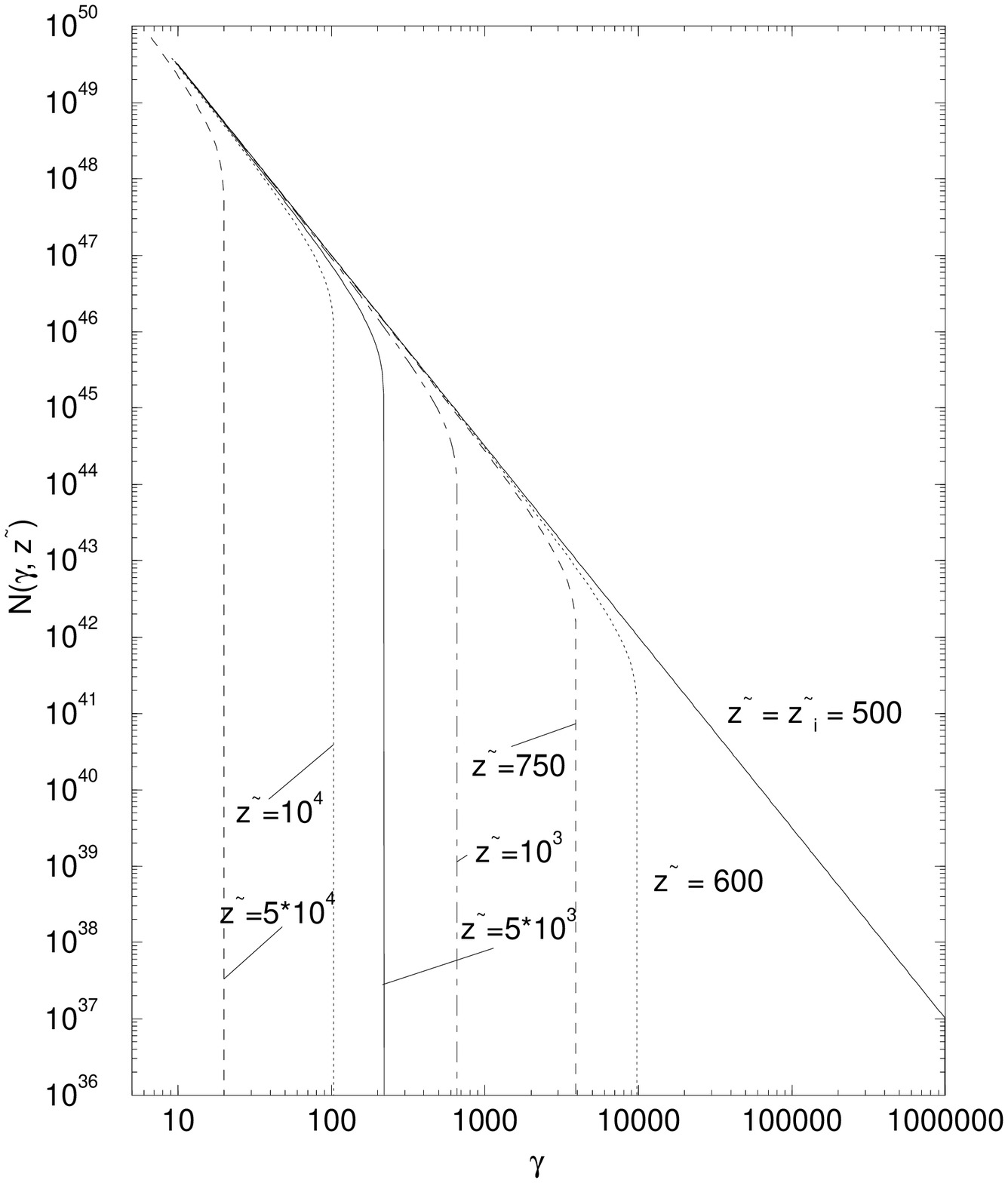} \abb
{\bf Fig. 7:  } Evolution of a power law particle energy spectrum
affected by synchrotron losses.  Parameters:  $\tilde z_i = 500$,
$\gepmi = 10$, $\gzpmi = 10^6$, $s = 2.5$, $M_8 = 1$, $B = 3 G$.
\bigskip \end{figure}

{}From Fig. 7 we see that the particle spectrum influenced by synchrotron
losses is well represented by a power law with the original spectral
index between the cutoffs given by Eq.  (25).  But in contrast to the
case of inverse Compton losses also the lower energy cutoff may be
considerably lowered and the higher energy cutoff does not approach a
fixed limit if the magnetic field strength remains constant along the
jet, as we assumed.

A rough estimate for the blob's distance $\tilde z_0$ from the accretion
disk plane at which the lower energy cutoff is significantly influenced
by synchrotron losses may be achieved by setting $\gepm (\tilde z_0) =
{\gepmi \over 2}$, yielding

$$ \tilde z_0^{sy} := \tilde z_i + {1 \over \gepmi \, K} \approx \tilde
z_i + 2 \cdot 10^7 {\Gamma_1 \over\gepmi \> M_8} \left( {B \over G}
\right)^{-2}.  \eqno(27) $$

\subsection{Bremsstrahlung losses}

An estimation of particle energy losses due to relativistic $e^+$-$e^-$
and $e^-$-$e^-$ bremsstrahlung is obtained using the formalism similar
to the one proposed by Dermer \& Liang (1989) for power law particle
spectra.  In contrast to their calculation, we include the Doppler
boosting of the bremsstrahlung photon assuming that it is emitted in the
direction of one of the incoming particles.  Furthermore, we use the
ultrarelativistic limit of the bremsstrahlung cross section derived by
Ba\u\i er et al.  (1967).

Results of this treatment are shown in Fig. 8 which shows that the
single particle energy loss rate is roughly proportional to the particle
energy so that the energy loss timescale remains roughly constant.

\begin{figure}
\epsfxsize=6cm
\epsffile[30 60 410 680] {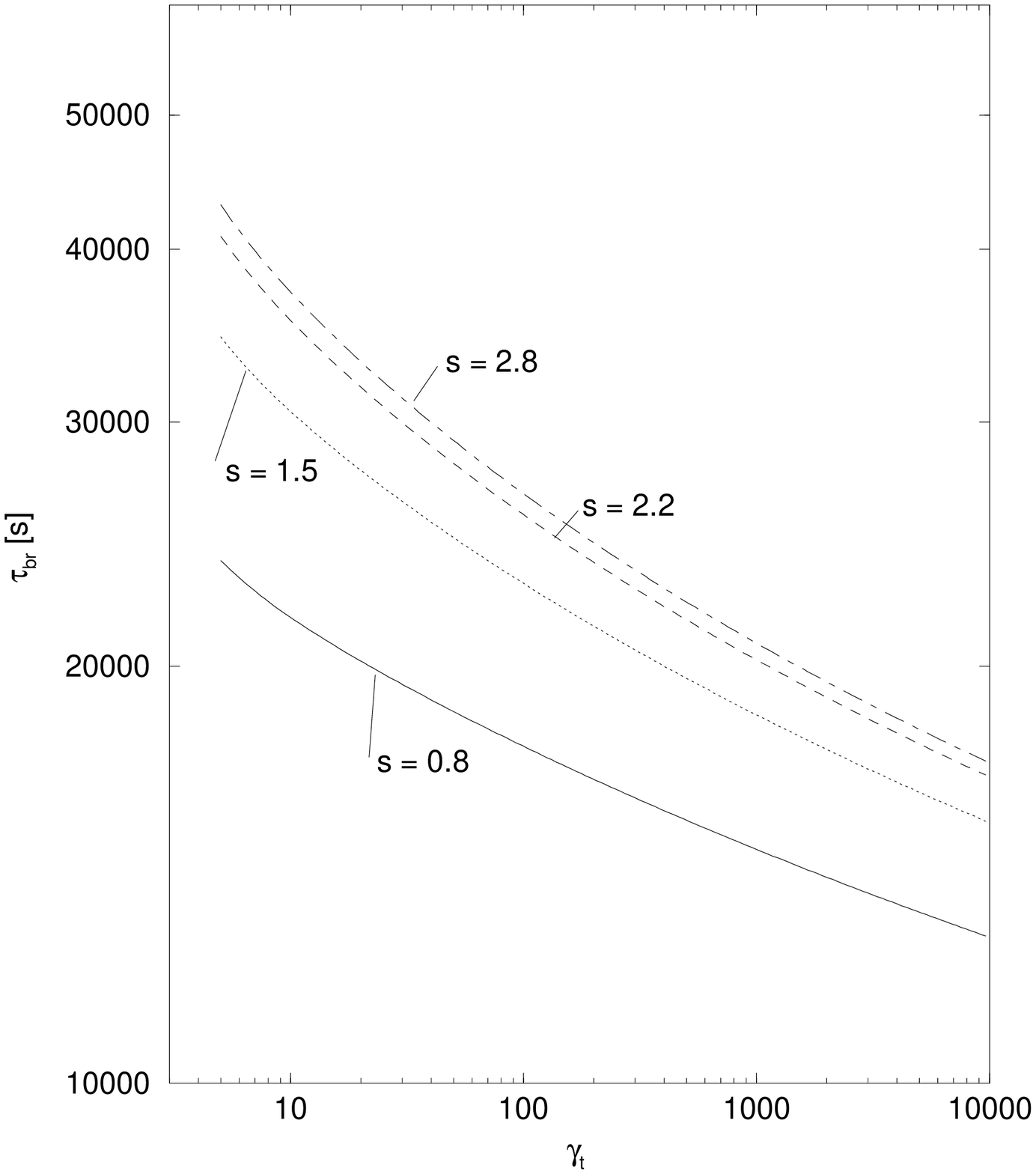}
\abb {\bf Fig. 8:  } Single particle energy loss timescales due to
bremsstrahlung emission for different values of the particle spectral
index; $n_{\pm} = 10^{11} \, cm^{-3}$.  \bigskip \end{figure}

As we also see from our numerical calculations, there is no very
significant dependence of the energy loss time scale on the lower
cutt-off of the particle distributions.  Considering the lower energy
end of the particle spectra, we find a typical energy loss timescale of
the order

$$ \tau_{br} = 5 \, \cdot \, 10^4 \, s \> \cdot \> n_{11}^{-1} \eqno(28)
$$

--- where $n_{11} = {n \over 10^{11} \, cm^{-3}}$ --- which corresponds
to a blob distance of

$$ \tilde z_0^{br} = \tilde z_i \, + \, 10^3 \> {\Gamma_1 \over n_{11}
\, M_8} \eqno(29).  $$

\subsection{Thermalization}

The thermalization timescale for a plasma described by a power law in
particle energy is again estimated using the formalism of Dermer \&
Liang (1989) to calculate the single particle energy loss and gain,
respectively, due to M\o ller and Bhabha scattering.  The numerical
results for M\o ller scattering for several values of the power law
spectral index $s$ are shown in Fig. 9; the diference between M\o ller
and Bhabha scattering turns out to be negligible.  We only consider the
timscales under the initial conditions; synchrotron, inverse Compton and
bremsstrahlung losses, of course, will alter the upper cut-off of the
particle distributions and hence, they will lengthen the thermalization
timescale for particles with lower energy.

\begin{figure}
\epsfxsize=6cm
\epsffile[30 30 410 750] {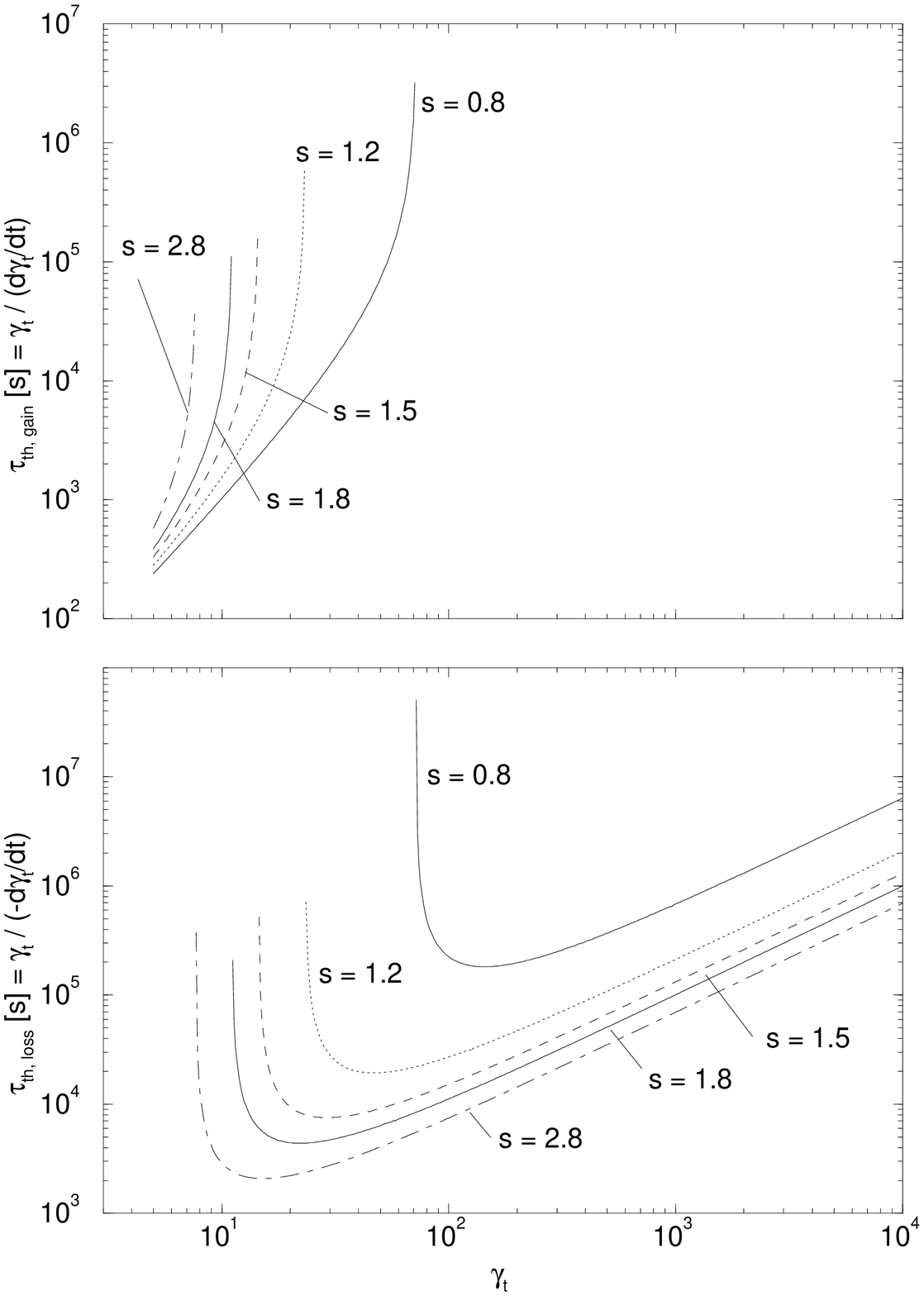}
\abb {\bf Fig 9: } Timescales for single particle
energy loss or gain, respectively due to M\o ller scattering; $n_{\pm} =
10^{11} \, cm^{-3}$ \bigskip \end{figure}

Obviously, the lower energy end of the particle spectrum is most
seriously affected by energy gain due to M\o ller scattering.  From our
numerical calculations we find the timescale for this energy exchange
process as

$$ \tau_{th} = 20 \, s \> {\gem \, \gep \over n_{11}} \eqno(30) $$

corresponding to a distance

$$ \tilde z_0^{th} = \tilde z_i \, + \, 0.4 \> {\Gamma_1 \, \gep \, \gem
\over n_{11} \, M_8}.  \eqno(31) $$

\subsection{Pair annihilation losses}

In the general case, the particle loss rate can be written as

$$ \left( {d \Npm (\gpm, \tilde z) \over d t} \right)_{PA} = -
\int_{V_B} d V \; n_{\pm} (\gamma_{\pm}, \tilde z) \; \cdot $$ $$ \cdot
\int\limits_{\gamma_{1\mp}}^{\gamma_{2\mp}} d\gmp \; n_{\mp}
(\gamma_{\mp}, \tilde z) \> \overline{ v \sigma } (\gamma_ +, \gamma_-)
\eqno(32) $$

which, in the ultrarelativistic limit and for a homogeneous blob,
becomes

$$ \left( {d \Npm (\gpm, \tilde z) \over d \tilde z} \right)_{PA} = -
\Npm (\gpm, \tilde z) {\lambda \pi r_e^2 \over \Gamma \, \bg \, \gpm}
\cdot $$ $$ \cdot \int\limits_{\gamma_{1\mp}}^{\gamma_{2\mp}} d\gmp \; {
n_{\mp} (\gmp, \tilde z) \over \gmp } \left( \ln[ 4 \gamma_+ \gamma_-] -
2 \right).  \eqno(33) $$

where $N^{\pm} (\gpm, \tilde z)$ is the spectral particle number and we
used Svensson's formula (19 b) for the angle averaged reaction rate
$\overline{v \sigma}$.

In order to estimate the timescale for the evolution of initial power
law particle spectra we may replace $n_{\mp} (\gmp, \tilde z)$ by its
initial value given by eq.  (1).  Then, the integral in eq.  (33) is
analytically solvable yielding

$$ \left( {d \Npm (\gpm, \tilde z) \over d \tilde z} \right)_{PA} = -
\Npm (\gpm, \tilde z) {P n_{\mp} (\gmp, \tilde z) \over \gemp \gpm}
\eqno(34) $$

where, for spectral indices $2 \le s \le 3$, $P$ may be roughly
estimated by

$$ P = {(s-1) \lambda \pi r_e^2 \over s \> \Gamma \, \bg } \, \biggl(
\ln [ 4 \gpm \gemp] + {1 \over s} - 2 \biggr) $$ $$ \approx \; 10^{-12}
\> {M_8 \over \Gamma} cm^3 \eqno(35) $$

\begin{figure}
\epsfxsize=6cm
\epsffile [30 60 410 680] {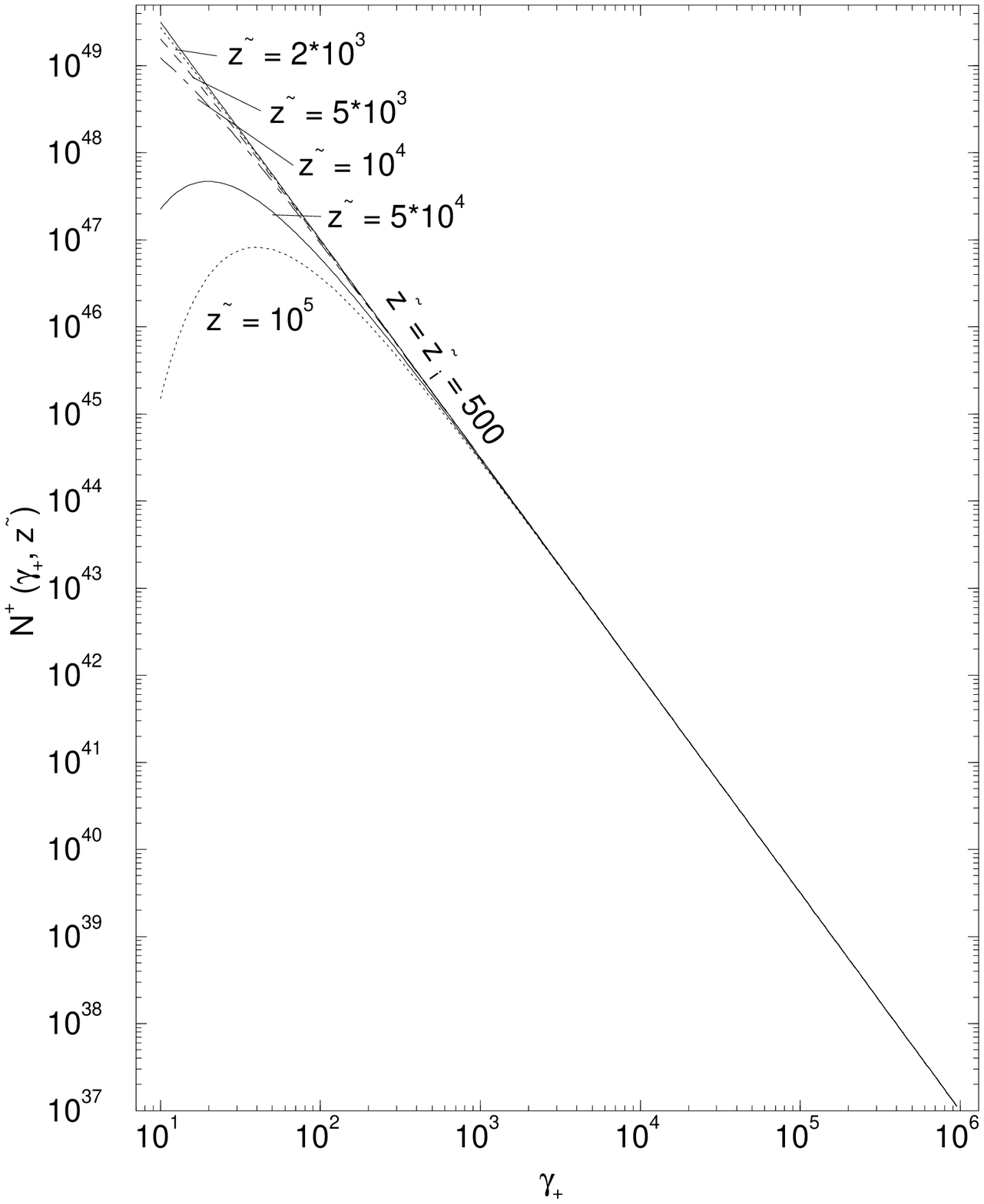} \abb
{\bf Fig. 10:  } Evolution of a power law particle energy spectrum
influenced exclusively by pair annihilation losses in the case of an
electron-dominated pair plasma.  Parameters:  $n_- = 10^{11} \, cm^{-3}
\, \gg \, n_+$, $\tilde z_i = 500$, $\gepmi = 10$, $\gzpmi = 1 0^6$, $s
= 2.5$, $M_8 = 1$, $\Gamma = 10$.  \bigskip \end{figure}

We now consider two special cases in which the differential equation
(34) is easily solvable.  First, we assume the pair plasma to be
dominated by electrons (an $e^-$-$p^+$-plasma), thus $n_- \gg n_+$.
Further neglecting geometrical effects on the extent of the plasma
blob, we set $n_- = const.$, and obtain

$$ N^+ (\gp, \tilde z) = N^+ (\gp, \tilde z_i) \> e^{- {n_- \, P \over
\gem \, \gp} \> (\tilde z - \tilde z_i)} \eqno(36) $$

which is shown in Fig. 10.  From Eq.  (36) we see the typical evolution
time scale to be

$$ \tilde z_0 = \tilde z_i + {1 \over A} \approx \tilde z_i + 10^{12}
{\Gamma \> \gem \> \gep \over M_8 \> n_- [cm^{-3}]}.  \eqno(37) $$

Secondly, we assume that electrons and positrons have the same particle
density and distribution functions and again neglect geometrical effects
eq.  (34) becomes

$$ \left( {d \Npm (\gpm, \tilde z) \over d \tilde z} \right)_{PA} = -
{\Npm}^2 (\gpm, \tilde z) {P \over V_b \> \gemp \gpm} \eqno(38) $$

where $V_b$ is the blob's volume.  This is solved by

$$ \Npm (\gpm, \tilde z) = {1 \over c_0 \> + \> {P \, \tilde z \over V_b
\, \gepm \gpm } }. \eqno(39) $$

where the constant $c_0$ can be estimated by ${1 \over \Npm (\gpm,
\tilde z_i)}$.  Thus we write

$$ \Npm (\gpm, \tilde z) = {\Npm (\gpm, \tilde z_i) \over 1 \, + \,
{\tilde z - \tilde z_i \over \tilde z_0 (\gpm)}}.  \eqno (40) $$

\begin{figure}
\epsfxsize=6cm
\epsffile [30 60 410 680] {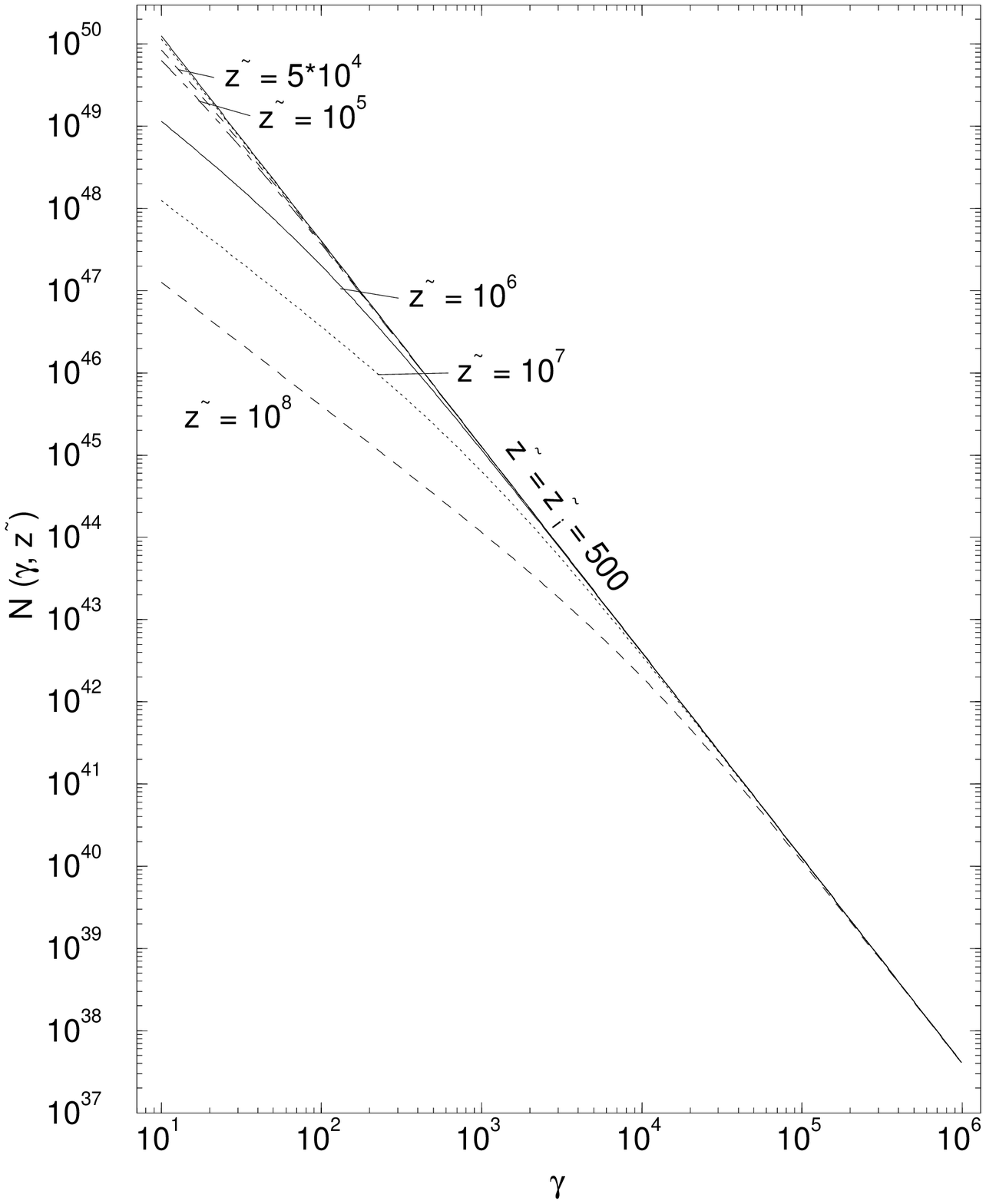} \abb
{\bf Fig. 11:  } Evolution of a power law particle energy spectrum
influenced by pair annihilation losses in the case of electrons and
positrons having the same density and distribution function.
Parameters:  $n_{\pm} = 10^{11} \, cm^{-3}$, $\tilde z_i = 50 0$,
$\gepmi = 10$, $\gzpmi = 10^6$, $s = 2.5$, $M_8 = 1$, $\Gamma = 10$.
\bigskip \end{figure}

This behavior which is, of course only self-consistent as long as the
particle spectra may be well fitted by a power law with spectral index
$\sim s$ is shown in Fig. 11.  We derive the typical evolution timescale
to be

$$ \tilde z_0 (\gpm) := {\gepm^3 \over P \> n_{\pm} (\tilde z_i)}
\approx 10^{12} \; {\gepm^3 \> \Gamma \over M_8 \> n_{\pm} (\tilde z_i)
[cm^{-3}]}.  \eqno(41) $$

where we used ${\Npm (\gpm, \tilde z_i) \over V_b} = n_{\pm} (\gpm,
\tilde z_i) \approx n_{\pm} {(s - 1) \, \gepm^{s - 1} \over \gpm^s}$.

\subsection{Comparison of timescales}

We now compare the typical timescales for particle spectra evolution by
the energy loss and exchange processes considered above.

First, we see that the typical energy loss timescale for the
inverse-Compton process is much shorter than for pair annihilation as
well as for all the other processes considered above, but it only
influences particles with energies above $\gu$.  Thus, it's negligible
in calculations to the yield of pair annihilation radiation.

For the other processes, we estimated that

$$ \tilde z_0 - \tilde z_i = \cases{ 10^2 \> {\gepm^3 \> \Gamma_1 \over
M_8 \> n_{11}} & for PA, $n_+ = n_-$ \cr 10^2 \> {\gep \gem \> \Gamma_1
\over M_8 \> n_{11}^-} & for PA, $n_+ \ll n_-$ \cr 10^7 {\Gamma_1 \over
\gepm \> M_8} \left({B \over G}\right)^{-2} & for synchrotron  \cr 10^3
\> {\Gamma_1 \over n_{11} \, M_8} & for bremsstrahlung \cr 0.4 \>
{\Gamma_1 \, \gep \, \gem \over n_{11} \, M_8} & for thermalization \cr
} \eqno(42) $$

Comparing these timescales we see that for lower cutoffs $\lte 3$ the
effect of bremsstrahlung losses are of minor importance to the particle
spectra evolution. Particularly for the lower energy end of the
particle spectrum, the bremsstrahlung losses are overestimated by
the cross section approximation used in our calculation. So, this
constraint will be eased when treating the bremsstrahlung emission
process in more detail. But because of being much more efficient than
bremsstrahlung emission energy gain due to M\o ller and Bhabha
scattering will compensate the according energy losses.

As can be seen in Fig. 9, thermalization will affect the lower energy
end of the particle spectra very seriously.  But since it is no energy
loss process and thus --- by its nature --- does not lead to a cooling
of the plasma the latter will remain ultrarelativistic. This energy
exchange process, of course, will alter the shape of the particle
spectrum at low energies, but the mean energy of the particles remains
constant, and the higher energy end of the particle spectrum will not
suffer heavy energy losses. Thus, for a detailed discussion of the
particle spectra evolution thermalization effects must be taken into
account, but they do not seriously alter the shape of the time
integrated annihilation spectrum.

Furthermore, we can derive a critical magnetic field strength for the
dominance of synchrotron losses over pair annihilation losses on
particle spectra evolution.  This is given by

$$ \left({B_{cr} \over G}\right)^2 = \cases{ 10^5 {n_{11} \over \gepm^4}
& for $n_+ = n_-$ \cr 10^5 {n_{11}^- \over \gep^2 \> \gem} & for
$n_+ \ll n_-$ \cr } \eqno(43) $$

If $B$ is significantly lower than $B_{cr}$ we may really neglect the
effect of synchrotron cooling on the particle energy spectrum.

\section{Comparison to the inverse-Compton spectrum}

We now compare the yield of pair annihilation photons to the
$\gamma$-ray spectrum produced by inverse-Compton scattering of
accretion disk photons by the high energy particles of the jet.  As we
pointed out in section 4.6, this process is only efficient for high
energy particles $\gamma > \gu$, but works over a much shorter timescale
than all the other processes occurring in the pair plasma. According to
the long integration times of the observing instruments we only see time
integrated spectra of AGN.

Thus, it is of interest to compare the time integrated inverse-Compton
spectrum to the pair annihilation spectrum integrated over the typical
timescale either for pair annihilation effects on the lower energy
particles or for any other mechanism reducing the particle density of
pairs near the lower energy cutoff.

\begin{figure}
\rotate[r]{
\epsfysize=6cm
\epsffile[100 40 600 500] {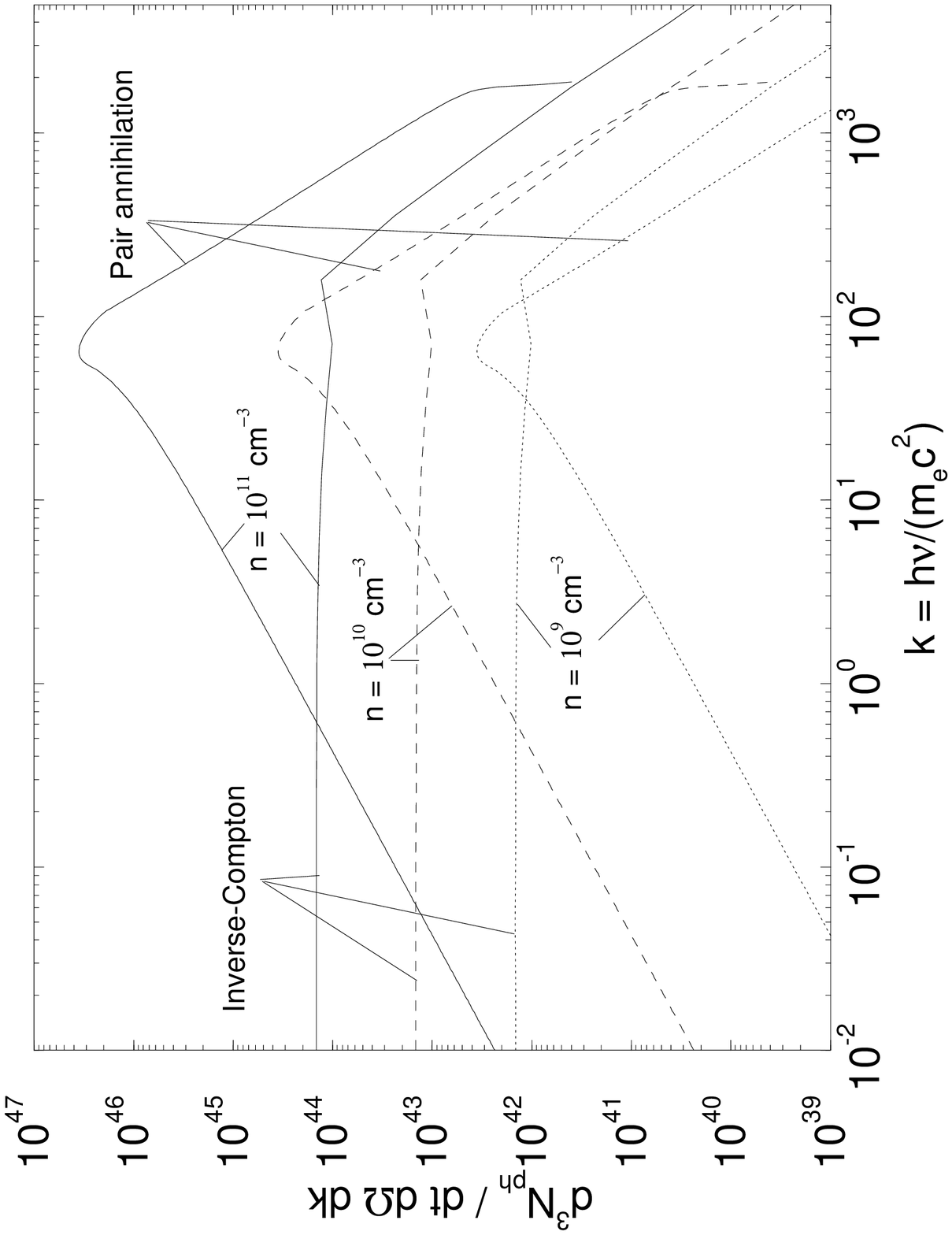} }
\abb {\bf Fig. 12:} Time-integrated inverse-Compton and pair
annihilation spectra; parameters in the text.
\bigskip \end{figure}

For the following calculations, we assume this timescale to be of the
order of $\tilde z_0 \sim 10^3$.  Each plasma blob is assumed to be
spherical with a radius of $10^{11} \, cm$ and injected at $\tilde z_i =
10$.  The accretion disk radiates with $L_0 = 10^{46} {erg \over s}$,
and the mass of the central black hole is $10^8 M_{\odot}$.
The accretion disk radiatiation spectrum is modelled by a greybody
distribution with a radial dependence of the tempearure according to the
accretion disk model of Payne and Eardley (1977).
The lower energy cutoff of the distribution functions is $\gepm = 3$,
the higher cutoff is $\gzpm = 10^4$, and we assume that both electron
and positron density are equal . The spectral index of the distribution
functions is $s = 2$.  The observing angle with respect to the jet axis
is $\theta = 2^o$, and the blob moves outwards with a Lorentz factor
of $\Gamma = 10$.  These parameters are convenient to explain the
$\gamma$- ray spectra of some $\gamma$-ray blazars by the
inverse-Compton process.

The inverse-Compton radiation spectrum is calculated using the full
Klein-Nishina cross section in the head-on approximation.
The resulting photon number spectra from inverse-Compton scattering and
pair annihilation are shown in Fig. 12 where we see that for particle
densities $n \gte 10^9 \, {\rm cm}^{-3}$ the contribution of pair
annihilation should lead to an observable effect in the spectra.
The flat inverse Compton spectrum at small photon energies is particularly
noteworthy.

We attribute this mainly to the presence of a low-energy cutoff at $E_c =
\gamma_c \, m c^2$ in the radiating particle distribution function. From the
work of Schlickeiser (1982) it is well known that such a low-energy cutoff
indeed leeds to a flat scattered photon spectrum below the photon energy
$$ E_{\gamma}^0 = {4 \, \epsilon_0 \gamma_c^2 \over 1 + {4 \epsilon_0 \gamma_c
\over m \, c^2}} \approx 4 \epsilon_0 \gamma_c^2, \eqno(44) $$
where $\epsilon_0$ denotes the energy of the assumed monoenergetic target
photon distribution function. (Corresponding calculations for extended target
photon energy distributions by Schlickeiser [1980] showed similar flattenings.)
According to Eqs. (14) and Fig. 1 of Schlickeiser (1982) the inverse Compton
power of scattered photons varies proportional to
$$ \chi(E_{\gamma}) \propto E_{\gamma}^{1 - s \over 2} \;\;\;\;\;\; {\rm for}
\;\; E_{\gamma} \ge E_{\gamma}^0 \eqno(45 a) $$
where $s$ is the power-law index above $E_c$, while at energies $\epsilon_0 \le
E_{\gamma} \ll E_{\gamma}^0$ one obtains
$$ \chi(E_{\gamma}) \propto \left( {E_{\gamma} \over \epsilon_0} \right)^{1 - s
\over 2} \left( {E_{\gamma} \over 4 \, \epsilon_0 \, \gamma_c^2} \right)^{s+1
\over 2} \propto \left( {E_{\gamma} \over \epsilon_0} \right), \eqno(45 b) $$
indicating that indeed spectral breaks in the inverse Compton scattered photon
spectrum by $\Delta \alpha= 1 - {1 - s \over 2} = {s + 1 \over 2}$ around the
intrinsic energy $E_{\gamma}^0$ are implied. Due to the blueshift by the
relativistically moving blob this spectral break occurs at
$$ k_c = {4 \, \epsilon_0 \, \gamma_c^2 \over m \, c^2} D (\mu^{\ast}) = 1.96
\, \left( {\epsilon_0 \over 10 \, keV} \right) \left( {\gamma_c \over 5}
\right)^2 D(\mu^{\ast}) \eqno(46) $$
with the Doppler factor
$$ D(\mu^{\ast}) = \Bigl[ \Gamma ( 1 - \bg \mu^{\ast}) \Bigr]^{-1}. \eqno(47)
$$
$k_c$ lies in the MeV energy band which is in agreement with the inverse
Compton curves in Fig. 12. With parameter values as chosen here, on top of this
break the contribution from the pair annihilation radiation is located.

\subsection{Comparison with observations}

It can be seen from Fig. 12,
that the pair annihilation emission in such a highly relativistic
plasma does not result in a 511-keV peak.  The annihilation radiation is
blueshifted both, by the relativistic motion of the pairs in the blob
rest frame, and by the Doppler boosting due to the bulk motion.
The pair annihilation peaks near the photon energy
$$k_{max}=\gamma _1D(\mu ^*)\eqno (48),$$
with the Doppler factor (47).

For the parameters chosen in Fig. 12 we obtain $k_{max}=89.2$, so that
the annihilation feature is at energies of $\sim 50$ --
$100 \, MeV$.  In this case the annihilation bump appears almost like the
well-known (e. g. Cavallo \& Gould 1971, Stecker 1973) $\gamma$-ray bump from
decaying $\pi^0$'s near $70 \, MeV$ if they do not move
relativistically.  The detection of enhanced bumpy emission near $70 \,
MeV$ from an AGN would therefore be no unique clue on the presence of
hadrons in the source, but may point to blueshifted annihilation
radiation.

However, for a sample of AGN sources seen under different viewing angles
$\mu^*$ there will be quite some variation in the values of the parameters
$\gamma _1$, defining the low-energy cutoff of the relativistic electron and
positron energy spectrum, and the relativistic bulk speed of the emitting blob,
$\Gamma$, from source to source. While the energy exchange processes discussed
in section 4, will keep the low-energy turnover of the distribution function at
mildly relativistic values, $\gamma _1\ge 3$, the wide dispersion in the
Doppler factor (47) values will imply a corresponding wide distribution in the
value of the annihilation peak energy (48). Although we have made no detailed
effort so far to optimize parameters to explain individual source spectra, it
is evident that in principle the inclusion of the annihilation radiation is
capable to explain emission maxima and spectral breaks at MeV energies.
The occurrence of spectral maxima between 1 and 10 MeV for the sources
3C273 and 3C279 has been reported by recent Comptel observations
(Williams et al. 1995). An even more drastic example is the detection of the
MeV-blazar GRO J0516-609 near PKS 0506-612/0522-611 by Bloemen et al. (1995).
Our spectral modelling confirms the suggestion of Bloemen et al. that the MeV
emission in this object represents the broad blue-shifted annihilation line.

A possible test for the hypothesis that the emission maxima in the observed
$\gamma $-ray energy spectra are due to blueshifted annihilation radiation is
offered by the predicted relation between the maximum photon energy (48) and
the total $\gamma $-ray luminosity. Making the strong assumption, that apart
from the Doppler factor all other intrinsic physical parameters of the sources
are equal, the total pair annihilation luminosity will scale with the Doppler
factor as $L_p\propto D^3$. A similar Doppler factor dependence holds for the
inverse-Compton luminosity (DS). Combined with Eq. (48) this implies the
relation
$$k_{max}\propto L_p^{1/3}\eqno (49),$$
i.e. the strong Doppler boosting in bright objects also blueshifts
the annihilation peak to high photon energies.

\begin{figure}
\epsfxsize=6cm
\rotate[r]{
\epsffile[85 25 600 200] {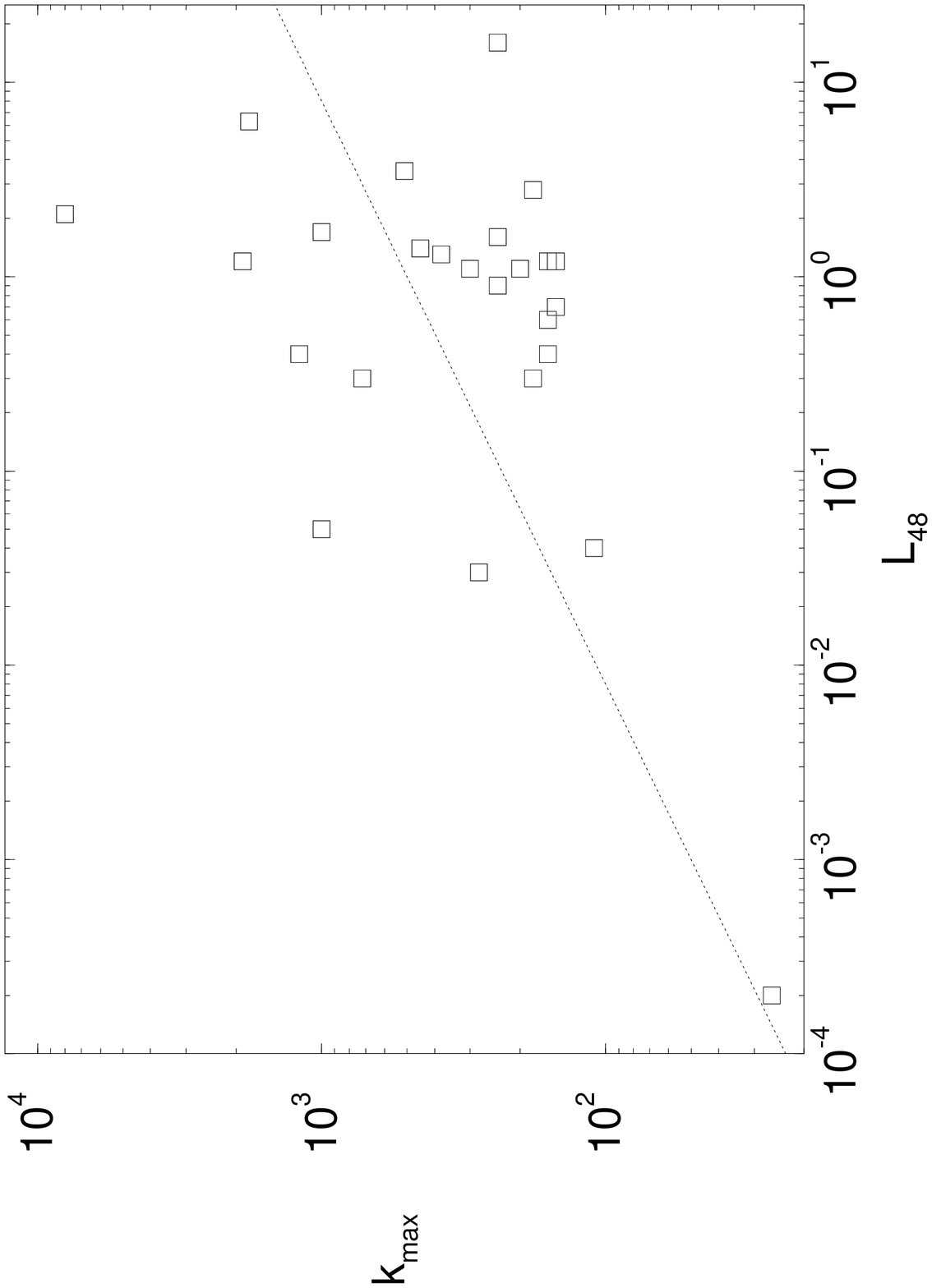} }
\abb {\bf Fig. 13:} Relation between MeV-$\gamma$-ray peak energy and total
$\gamma$-ray luminosity for 33 AGN detected by EGRET. \bigskip
\end{figure}

We have used the available compilation of blazar energy spectra
(Montigny et al. 1995) to estimate both the $\gamma$-ray luminosity $L_p$
and the maximum photon energy (correcting for cosmological redshift),
and the result is shown in Fig. 13 in comparison with the predicted
correlation (49). The presently available data are not in contradiction
with the hypothesis (49). A more stringent test would be possible if more
AGNs with luminosities in the range $0.0005-0.03L_{48}$ are detected
with well-resolved $\gamma $-ray energy spectra.

\section{Summary and conclusions}

In this work we have calculated the contribution of the pair
annihilation process in relativistic electron-positron jets to the
$\gamma -$ray emission.  Under the same assumptions as for the
calculation of the yield of the inverse Compton scattered photons by
relativistic particles in the jet (Dermer and Schlickeiser 1993) we
calculated the emerging pair annihilation radiation taking into account
all spectral broadening effects due to the energy spectra of the
annihilating particles and the bulk motion of the jet.  We started from
the pair annihilation reaction rate in relativistic plasmas (Svenson
1982) and developed a new asym ptotic approximation appropriate for
relativistic pair energies which provides a very accurate approximation
of the pair annihilation spectrum for all photon energies.

The evolution of the particle energy spectra with time has been taken
into account consistingly, discussing in detail the role of inverse
Compton losses, synchrotron losses, relativistic bremsstrahlung losses,
pair annihilation losses and M\o ller and Bhabha thermalization. For
typical blazar parameters two particle-number conserving loss processes
dominate at large (Lorentz factor $\gamma >10^3$) particle energies
(inverse Compton losses) and at very low (Lorentz factor $\gamma <30$)
energies (thermalization), while the catastropic pair annihilation los
ses reduce the number of radiating particles thus avoiding the
"dead-electron-problem" (Begelman et al. 1984).  We demonstrated that in
spectral appearance the time-integrated pair annihilation spectrum
appears almost like the well-known $\gamma $-ray spectrum from decaying
$\pi^o$-mesons at rest, since it yields a broad bumpy feature located
between 50 and 100 MeV. Moreover, we demonstrated that for pair
densities greater $10^{10}$ cm$^{-3}$ in the jet the annihilation
radiation will dominate the inverse Compton radiation, and due to its
bump-like spectrum indeed may explain reported spectral
bumps at MeV energies. The refined treatment of the inverse Compton
radiation including Klein-Nishina cross section corrections,
the low-energy cutoff in the radiating relativistic particle distribution
function and a realistic greybody accretion disk target photon spectrum
leads to spectral breaks of the inverse Compton emission in the MeV energy
range with a change in spectral index $\Delta \alpha$ larger than 0.5
as detected in PKS~0528+134 and 3C273.

\noindent {\it Acknowledgements} We thank H. Mause for calculating the inverse
Compton radiation
spectra.
Discussions with C. D. Dermer, F. Aharonian, M. Pohl and G. Henri are
gratefully acknowledged. RS acknowledges partial support by the DARA
(50 OR 94063) of his Compton observatory guest investigator programs. MB
acknoledges financial support by the Deutsche Forschungsgemeinschaft.


\begin{thebibliography}{}

\bibitem[1968]{1} Ba\u\i er, V. N., Fadin, V. S., Khoze, V. A., 1967, Soviet
Physics JETP {\bf 24}, Nr. 4, 760

\bibitem[1984]{2} Begelmann, M. C., Blandford, R. D., Rees, M., 1984,
Rev.  Mod. Phys.  {\bf 56}, 255

\bibitem[1990]{3} Blandford, R. D., 1990, in:  Active Galactic Nuclei
(eds.  T. Courvoisier, M. Mayor), Springer, New York, p. 161

\bibitem[1993]{4} Blandford, R. D., 1993, in:  Proc. of the Compton Symp.
(eds.  N. Gehrels, M. Friedlander, \& J. P. Norris), AIP, New York, p.
533

\bibitem[1979]{5} Blandford, R. D., K\"onigl, A., 1979, ApJ {\bf 232}, 34


\bibitem[1979]{5} Bloemen, H., Bennett, K., Blom, J. J., et al., 1995, AA {\bf
293}, L1

\bibitem[1971]{6} Cavallo, G., Gould, R. J., 1971, Nuovo Cimento {\bf 2
B}, 77

\bibitem[1990]{7} Coppi, P. S., Blandford, R. D., 1990, MNRAS {\bf 245},
453

\bibitem[1993]{8} Dermer, C. D., 1993, in:  Proc. of the Compton Symp.
(eds.  N. Gehrels, M. Friedlander, \& J. P. Norris), AIP, New York, p.
541

\bibitem[1994]{9} Dermer, C. D., Gehrels, N. 1995, ApJ,in press

\bibitem[1989]{10} Dermer, C. D., Liang, E. P., 1989, ApJ {\bf 339}, 512

\bibitem[1992]{11} Dermer, C. D., Schlickeiser, R., 1992, Science {\bf
257}, 1642

\bibitem[1993]{12} Dermer, C. D., Schlickeiser, R., 1993, ApJ {\bf 416},
458 (DS)

\bibitem[1994]{13} Dermer, C. D., Schlickeiser, R., 1994, ApJ Suppl.
{\bf 90}, 945

\bibitem[1993]{14} Henri, G., Pelletier, G., Roland, J., 1993, ApJ {\bf
404}, L41

\bibitem[1994]{15} Kanbach, G., 1994, private communication

\bibitem[1993]{16} Kurfess, J. D., 1994, in:  Multi-wavelength continuum
emission of AGN, eds.  T.J.-L.  Courvoisier, A. Blecha, Kluwer,
Dordrecht, p. 39

\bibitem[1987]{17} Lightman, A. P., Zdziarski, A. A., 1987, ApJ {\bf
319}, 643

\bibitem[1995]{17} v. Mongitny, C., Bertsch, D. L., Chiang, J., et al., 1995,
ApJ, in press

\bibitem[1994]{18} Payne, D. G., Eardley, D. M., 1977, Ap Lett.
{\bf 19}, 39

\bibitem[1992]{19} Protheroe, R. J., Mastichiadis, A., Dermer, C. D.
1992, Astroparticle physics {\bf 1}, 113

\bibitem[1979]{20} Scheuer, P.A.G., Readhead, A.C.S., 1979, Nature {\bf
277}, 182

\bibitem[1981]{20a} Schlickeiser, R., 1980, ApJ {\bf 240}, 636

\bibitem[1982]{20b} Schlickeiser, R., 1982, A \& A {\bf 107}, 378

\bibitem[1982]{21} Svensson, R., 1982, ApJ {\bf 258}, 321

\bibitem[1973]{22} Stecker, F. W., 1973, ApJ {\bf 185}, 499

\bibitem[1994]{23} Thompson, D., Bertsch, D. L., Dingus, B. L., et al., 1994,
in:  Multi-wavelength continuum emission of AGN, eds.  T.J.-L.  Courvoisier,
A. Blecha, Kluwer, Dordrecht, p. 49

\bibitem[1976]{24} Williams, O. R., Bennett, K., Bloemen, H., et al. 1995, AA
{\bf }, in press

\bibitem[1976]{25} Zdziarski, A. A., Ghisellini, G., George, I. M., et al.
1990, ApJ {\bf 363}, L1

\end{thebibliography}
 \end{document}